# Emergent ferromagnetism with Fermi-liquid behavior in proton intercalated CaRuO$_3$

*Shengchun Shen[1], Zhuolu Li[1], Zijun Tian[2], Weidong Luo[2,3*], Satoshi Okamoto[4*] and Pu Yu[1,5,6*]*

[1]*State Key Laboratory of Low Dimensional Quantum Physics and Department of Physics, Tsinghua University, Beijing 100084, China*

[2]*Key Laboratory of Artificial Structures and Quantum Control, School of Physics and Astronomy, Shanghai Jiao Tong University, Shanghai 200240, China*

[3]*Institute of Natural Sciences, Shanghai Jiao Tong University, Shanghai 200240, China*

[4]*Materials Science and Technology Division, Oak Ridge National Laboratory, Oak Ridge, Tennessee 37831, USA*

[5]*Frontier Science Center for Quantum Information, Beijing 100084, China*

[6]*RIKEN Center for Emergent Matter Science (CEMS), Wako 351-198, Japan*

Email: wdluo@sjtu.edu.cn; okapon@ornl.gov and yupu@mail.tsinghua.edu.cn

**The evolution between Fermi liquid and non-Fermi-liquid states in correlated electron systems has been a central subject in condensed matter physics because of the coupled intriguing magnetic and electronic states. An effective pathway to explore the nature of non-Fermi liquid behavior is to approach its phase boundary. Here we report a crossover from non-Fermi liquid to Fermi-liquid state in metallic CaRuO$_3$ through ionic liquid gating induced protonation with electric field. This electronic transition subsequently triggers a reversible magnetic transition with the emergence of an exotic ferromagnetic state from this paramagnetic compound. Our theoretical analysis reveals that hydrogen incorporation plays a critical role in both the electronic and magnetic phase transitions via structural distortion and electron doping. These observations not only help understand the correlated magnetic and electronic transitions in perovskite ruthenate systems, but also provide novel pathways to design electronic phases in correlated materials.**

## I. INTRODUCTION

Landau Fermi liquid (FL) theory has been a successful model for describing the low-temperature electronic properties of metals [1]. However, recently there have appeared an increasing number of systems that challenge the concept of FL by showing anomalous properties with non-Fermi liquid (NFL) behavior [2-8]. How to understand the NFL behavior forms an essential topic in the study of strongly correlated systems. Among these systems, transition metal oxides have attracted particular interest due to their vastly unusual properties with notable NFL behavior [9-14], and the corresponding investigation is an important factor toward understanding of their fascinating properties. For instance, near a magnetic transition, the associated order parameter fluctuations may lead to the breakdown of FL and emergence of NFL behavior [4,5,15]. Although extensive investigations have been carried out in heavy fermion compounds and metallic alloys [4-6], the correlation of NFL and magnetic phase transition in transition metal oxides remains rare [11,12,16], which is likely due to the lack of effective pathways to obtain delicate manipulation of their magnetic and electronic states across these transitions.

Ruthenates represent an important family of complex oxides with rich spectrum of properties, ranging from unconventional superconductivity to ferromagnetism [12,17-19]. Among this family, $CaRuO_3$ has been extensively studied due to its unique non-magnetic metallic state, as well as notable NFL behavior [16,20-22]. Interestingly, its isoelectronic $SrRuO_3$ shows an ferromagnetic ground state with FL behavior [19], and such distinct magnetic states were recognized due to the change of crystalline structure [23,24]. The contrasting magnetic and electronic properties between $SrRuO_3$ and $CaRuO_3$ suggest an intimate correlation between the Fermi-liquid behavior and ferromagnetic state in those systems. Hence, $CaRuO_3$ provides a great playground to explore the correlation between NFL behavior and magnetic phase transition if one can introduce the ferromagnetic state into this system. Chemical substitution with B site isovalent chemical ions was widely employed to manipulate the magnetic state in

CaRuO$_3$ in the last several decades with the aim to probe the magnetic and electronic transition simultaneously [25-31], which however generally leads to electron localization rather than itinerant ferromagnetism, and the resulted magnetism was likely attributed to the suppressed long range magnetic ordering with local cluster [32-34]. While Sr$_{1-x}$Ca$_x$RuO$_3$ system undergoes a transition from ferromagnetic metal with FL behavior (FM-FL) to paramagnetic metal with NFL behavior (PM-NFL) with the increase of Ca concentration, the details close to the phase boundary differ substantially among previous reports limited by sample-to-sample variation [35-38]. Besides, previous theoretical work suggested that an itinerant ferromagnetic state can be potentially obtained from CaRuO$_3$ through the epitaxial tensile strain [39], while the experimental results are controversial [36,40]. Therefore, it remains elusive whether a ferromagnetic FL metallic state could be obtained from CaRuO$_3$, and if achievable how its FL behavior is correlated with the ferromagnetic state near the transition.

It is interesting to note that a recent study revealed a convenient method to control the magnetic state of SrRuO$_3$ through the ionic liquid gating (ILG) induced proton intercalation (protonation) with the associated electron doping and structural deformation, which leads to a ferromagnetic to paramagnetic transition, while maintaining its robust FL behavior at low temperatures [41]. This study evokes immediately a series of interesting questions: whether the protonation process can lead to a novel magnetic state in CaRuO$_3$, and how it would be coupled with the NFL behavior? In this letter, we demonstrate a reversible protonation of pristine CaRuO$_3$ thin films through ILG. As a consequence, we observe an emergent ferromagnetic state as evidenced by the anomalous hall effect. Through careful analysis of the temperature-dependent transport data, we reveal an interesting NFL to FL crossover as the precursor of the magnetic transition. We attribute this distinct PM-NFL to FM-FL transition to the protonation induced band structural modulation as well as suppressed electronic correlation.

## II. RESULTS

## A. Proton intercalation in CaRuO₃ film

High quality thin films of CaRuO$_3$ were grown on (001) (LaAlO$_3$)$_{0.3}$-(SrAl$_{0.5}$Ta$_{0.5}$O$_3$)$_{0.7}$ (LSAT) substrates by a pulsed laser deposition system. Our samples show fully coherent epitaxial nature due to the relatively small lattice mismatch between sample and substrate. Figure 1(a) shows a schematic drawing of the ILG induced proton intercalation into the sample, in which protons are produced through the electrolysis of residual water within the ionic liquid (DEME-TFSI) and then inserted into the sample with the application of positive voltage across the gating electrode and sample [41-43]. To maintain charge neutrality, electrons are injected from the counter electrode into sample, completing the protonation process. With this setup, we first carried out *in-situ* X-ray diffraction (XRD) measurements during the ILG of CaRuO$_3$ thin films. As shown in Fig. 1(b), with increasing gate voltage ($V_G$), we observe a clear shift of the CaRuO$_3$ (001) diffraction peak toward lower angles when $V_G$ is larger than 1.5 V, and such a critical voltage is consistent with the previous study of SrRuO$_3$, and is attributed to the water electrolysis process [41]. The *c* lattice constant changes gradually from the pristine state and is eventually saturated with expansion of ~4.1%, providing the opportunity to manipulate continuously the electronic state of CaRuO$_3$. It is important to note that the in-plane epitaxial strain remains unchanged throughout the phase transformation as evidenced by *in-situ* RSM measurements (Fig. S1 [44]). Similar to the previous study of SrRuO$_3$, the structural transformation is also volatile, i.e., the crystal structure returns back to its pristine state when the gate voltage is turned off. The slight offset of the XRD peak (as shown in Fig. S1(a) [44]) is attributed to small amount of residual hydrogen within the sample, as supported by our *ex-situ* SIMS measurements on gated samples (Fig. 1(c)), in which about 10% hydrogen is observed in the post-gated H$_x$CaRuO$_3$.

## B. Proton intercalation induced magnetic phase transition

To trace the evolution of magnetic and electronic states in CaRuO$_3$ thin films through protonation, we performed *in-situ* transport measurements during ILG (see methods in

Supplementary Information [44]). Figure 2(a) shows the temperature dependent resistivity $\rho_{xx}(T)$ at different $V_G$. At lower $V_G$, the resistivity remains almost unchanged, while when $V_G$ is above 1.5 V, the resistivity is slightly enhanced, which is attributed to the local lattice distortion induced by the protonation. With the observation of robust metallic state, we then study its magnetic state through the anomalous Hall effect (AHE) measurements [45]. For pristine $CaRuO_3$ film, the Hall resistivity shows a linear dependence on the magnetic field (Fig. 2(b)), indicating a non-magnetic state. Surprisingly, the gated samples (e.g. $V_G$ = 2.5 V) exhibits a distinct AHE hysteresis loop with nonzero remnant Hall resistivity at zero magnetic field, which signals an emergent ferromagnetic state through the ILG. Besides, the magnetic field angle dependence of Hall resistance and longitudinal resistance suggests the magnetic easy axis in ferromagnetic $H_xCaRuO_3$ is along out-of-plane direction (Fig. S3). It is important to note that the emergence of ferromagnetism by the protonation is exactly opposite to that of $SrRuO_3$, in which the AHE, i.e. ferromagnetism, is totally suppressed through the protonation [41]. Furthermore, although the well-defined AHE hysteresis signal an emergent remnant ferromagnetization in proton intercalated $CaRuO_3$, its actual spin configuration remains undetermined and left for future investigations.

To trace the evolution of the ferromagnetic state in $H_xCaRuO_3$, we then systematically measured the temperature dependent Hall resistivity as a function of $V_G$ (Fig. S2 [44]). Figure 2(c) summarizes the temperature dependent AHE resistivity $\rho_{XY}^{AHE}$ at different $V_G$, in which $\rho_{XY}^{AHE}$ is defined as the saturated AHE component as shown in the inset of Fig.2(c). The systematic measurements reveal that the AHE (i.e. ferromagnetism) is observed with $V_G$ above 1.5 V, with the maximum transition temperature around 20 K. More importantly, the $V_G$ dependence of AHE is nonmonotonic, exhibiting a dome shape with increasing $V_G$ (Fig. 2(d)). Furthermore, the extracted carrier concentration (hole type) increases by almost one order of magnitude at initial stage of ILG and then decreases gradually for higher $V_G$ (Fig. 2(d)). The notable similarity between the $V_G$ dependent carrier concentration and anomalous Hall resistivity suggests the emergence of ferromagnetism in $H_xCaRuO_3$ is correlated with the electronic state. The discrepancy between the increase of hole concentration and hydrogen intercalation induced electron doping indicates that the Fermi surface of sample is composed of both electron-like and hole-like pockets.

**C. Proton intercalation induced NFL to FL transition**

With this established magnetic phase diagram, we then further investigate the evolution of NFL through the ILG. Generally, NFL manifests itself into different power-law behavior of physical quantities from those of a FL. For instance, the temperature dependent resistivity can be fitted with $\rho \sim T^\alpha$, where $\alpha = 2$ for FL and typically $\alpha < 2$ for NFL [2,3,5,16,20]. To explore the potential variance in electronic behavior of the ILG sample, we quantitatively analyze the diagonal resistivity $\rho_{xx}(T)$ by fitting the curves with the empirical relation $\rho \sim T^\alpha$ at low temperatures (Fig. S4 [44]). Intriguingly, the exponent $\alpha$ clearly shows two regimes. As shown in Fig. 3(a), the pristine film and gated samples with $V_G$ of 0.5 and 1.0 V show an obvious power law behavior with $\alpha \sim 3/2$ at low temperatures, and such a NFL behavior is typically attributed to the diffusive electron motion induced by strong interactions between itinerant electrons and the critically damped long-wavelength magnons in quantum critical point systems [20,46]. With increasing $V_G$ at 1.5 V and above, $\alpha$ changes from ~3/2 to ~2 (Fig. 3(b)), indicating an electronic transition from a NFL to a FL. Noteworthily, the Hall measurements reveal that the ferromagnetic state (with $\rho_{XY}^{AHE} \neq 0$) emerges only at $V_G > 1.5$ V (Figs. 3(c) and S2 [44]), which clearly indicates that the ferromagnetism emerges subsequently after the FL, rather than emerging simultaneously. This further suggests that the FL appears as a precursor of the ferromagnetism in CaRuO$_3$ system, which is consistent with the fact that the FL is robust through the magnetic transition in SrRuO$_3$.[39] Furthermore, it is also found that this magnetic and electronic transition can be induced reversibly with the application of gating voltage, showing an exotic magnetoelectric coupling. More specifically, when $V_G$ is cycled between 0 V and 3.5 V, the ON/OFF switching of ferromagnetic states can be realized, and the electronic behavior translates between NFL and FL as well, as shown in Figs. 3(d) and S5 [44].

## D. Mechanism of proton intercalation induced phase transitions

To gain insight into the emergent ferromagnetism and FL behavior in $H_xCaRuO_3$, we carried out density functional theory (DFT) calculations, and then integrated them with dynamical mean-field theory (DMFT) calculations, i.e., DFT+DMFT. The details are presented in Supplementary Information [44]. The optimized crystalline structure of $H_xCaRuO_3$ (x = 0.5) is shown in the Fig. 1(a), in which the intercalated proton ions prefer to form bonding with equatorial oxygen ions in $RuO_6$ octahedra. A systematic theoretical calculation reveals that along the protonation, the crystalline structure process an enlarged Ru-O-Ru bonding angle and expanded crystalline lattice with larger c/a ratio (tetragonality) (Fig. S6 [44]). Consequently, such structure modification results in the splitting of degenerate Ru $t_{2g}$ bands as well as narrower bandwidth (Fig. S6 [44]). Furthermore, with increasing intercalated proton concentration, the electron band spectra weight shifts gradually toward lower energy, due to the electron band filling, in which the $Ru^{4+}$ turns into $Ru^{3+}$ for the fully protonated case. More importantly, the proton intercalation leads to the characteristic modulation of density of states (DOSs) at the Fermi level, as revealed in Figs. S7 and S8 with DFT and DFT+DMFT methods respectively [44]. The pristine $CaRuO_3$ shows suppressed spectral weights at Fermi level ($E = 0$) (Fig. S7(a) and S8(a) [44]), which is consistent with earlier theoretical study [23] and account for the paramagnetic ground state following the Stoner criteria. By contrast, the proton intercalation process leads to a pronounced enhancement of DOSs in the van Hove singularity around the Fermi level (Figs. S7(b-d) and S8(b-d) [44]). Such peaked DOS could potentially have the Stoner instability, leading to ferromagnetic ordering in itinerant electron system [47,48]. As expected, ferromagnetism is found to be stabilized in DFT+DMFT for some $H_xCaRuO_3$ cases, e.g., $H_{0.5}CaRuO_3$ in Fig.4a (Fig. S9 [44]). Relatively small difference between the minority spin and majority spin DOS suggests that $H_{0.5}CaRuO_3$ is a weak ferromagnet, which is consistent with our experimental observations of low Curie temperature and small AHE. In addition, the electronic behavior also changes with proton intercalation, as indicated by the imaginary part of the self-energy (Figs. 4(b) and S8 [44]). For pristine CaRuO$_3$, it behaves as $(\omega_n)^{1/2}$, which indicates a non-Fermi-liquid behavior due to the spin freezing (Fig. 4(b)) [22]. For H$_{0.5}$CaRuO$_3$, the frequency range of this $(\omega_n)^{1/2}$ behavior becomes narrower and the absolute value of the self-energy becomes much smaller than that of CaRuO$_3$, suggesting it is more Fermi liquid like (Fig. 4(b)). In accordance with our experimental observations, DFT+DMFT calculations also manifest that proton intercalation induces a magnetic phase transition following the evolution from NFL to FL behavior.

As mentioned above, the proton intercalation process could have two contributions to influence the magnetic and electronic states: i) modifying band structure through lattice distortion and ii) doping electrons. Therefore, to quantitatively trace the ferromagnetism caused by proton intercalation, we carried out DFT+DMFT calculations by changing the band structure for H$_x$CaRuO$_3$ and the Ru $d$ electron number $N_e$ independently. One can see that $M$ is nonzero and changes nonmonotonically with varying $N_e$ at fixed $x$ (Fig. 4(c)), suggesting that the ferromagnetism is correlated with the electron number $N_e$. $M$ also changes dramatically by simply considering modified band structure induced by different hydrogen concentrations with fixed $N_e$=4, indicating ferromagnetism is also corelated with the structural distortion. The inset of Fig. 4(c) shows a resulting theoretical phase diagram of H$_x$CaRuO$_3$ system as a function of hydrogen concentration $x$ and electron number $N_e$. Ferromagnetic ordering is pronounced at the intermediate $x$ and $N_e$ as indicated by a yellow line, accompanied with Fermi liquid behavior emerging from a paramagnetic non-Fermi liquid state. In the actual experimental situation, the electron number depends on the hydrogen concentration as $N_e = 4 + x$. As shown as a dashed line in the inset of Fig. 4(c), this $N_e$ goes across the ferromagnetic regime extending from the upper left corner to the lower right corner, resulting from the combination of structural distortion, which pushes down the position of the van Hove singularity, and electron doping, which pushes up the Fermi level. Thus, the theoretical phase diagram is qualitatively consistent with the dome-shape AHE signal we observed. It is interesting to note that the largest $M$ is found at $N_e = 4$ with the fully proton intercalated HCaRuO$_3$ structure (Fig. S10 [44]). This suggests that stronger ferromagnetism could be achieved by doping holes into the fully proton intercalated HCaRuO$_3$ or by inducing the similar structural distortion as HCaRuO$_3$ without changing the nominal Ru valence by, for example, He implantation [49].

## III. DISCUSSION

Due to the fact that the 4d transition metal ions typical process more extended $d$-orbitals, and therefore the $p$-$d$ hybridization of O-Ru bond is more sensitive to the lattice distortion. Therefore, the subtle balance and competition between strong $p$-$d$ hybridization, electron-lattice coupling as well as the Coulomb interaction in $4d$ transition metal oxides can lead to dramatic evolution of the corresponding electronic and magnetic states. Our current results strongly suggest that the intercalated hydrogen atoms can readily tip the balance between different phases, leading to drastic changes in magnetic and electronic properties. As shown in above observations, the protonation in CaRuO$_3$ leads to a magnetic phase transition, which is correlated with both lattice distortion and electron doping induced by intercalated protons. Additionally, both experimental and theoretical results suggest that the Fermi liquid transition forms a precursor of the induced ferromagnetism in proton intercalated CaRuO$_3$ system. It is interesting to note that our previous work on proton intercalated SrRuO$_3$ suggests the FL behavior is maintained throughout the FM to PM magnetic transition [41]. These results are clearly contrary to a general consensus of FL instability at magnetic phase transition [5], in which the ferromagnetism and FL behavior is decoupled in Ca(Sr)RuO$_3$ systems. The NFL behavior is driven by strong electronic correlations that freeze local spin moments [22], the protonation could weaken the electronic correlations by changing $d$ electron density away from an integer value 4. It is clear that the FL behavior represents the dominant factor for the emergent ferromagnetism in proton intercalated CaRuO$_3$, due to the fact that the electron correlation modified by protonation plays a critical role for the magnetic transition. The whole process of proton intercalation CaRuO$_3$ could be understood as follows: The protonation process first leads to the breakdown of NFL in H$_x$CaRuO$_3$ with relatively small amount of H concentration due to the suppressed electron correlation. Then with further protonation, a ferromagnetic ground state emerges due to Stoner instability caused by modified band structure and suppressed electron correlation. While for weakly correlated SrRuO$_3$ system, the FL behavior remains in spite of an induced FM to PM transition through protonation.

In summary, we demonstrated that an itinerant ferromagnetic state can be introduced in paramagnetic CaRuO$_3$ via ILG induced protonation. As a precursor of this magnetic phase transition, the NFL to FL transition in electronic behavior appears due to the suppressed electronic correlation. More importantly, these transitions can be manipulated reversibly in an electric manner, suggesting an exotic magnetoelectric coupling effect. Our work not only leads to a further understanding of the itinerant ferromagnetism as well as NFL and FL behaviors in (Ca, Sr)RuO$_3$ systems, but also highlights that the electrically controllable protonation could serve as an effective pathway to manipulate the electronic and magnetic phase transitions in strongly correlated electron systems. It could trigger the general research interest to design the phase diagram of ruthenate systems, and a rich spectrum of novel properties could emerge.

## ACKNOWLEDGEMENTS

This work was financially supported by the Basic Science Center Program of NSFC (grant No. 51788104); NSFC (grant Nos. 51872155, U1632272, 11521404, 11904196, and 52025024); the National Basic Research Program of China (grant No. 2016YFA0301004); the Beijing Natural Science Foundation (Grant No. Z200007); the Tsinghua University Initial Science Research Program (20203080003) and the Beijing Advanced Innovation Center for Future Chip (ICFC); and the Engineering and Physical Sciences Research Council (grant EP/N016718/1). The research by S.O. was supported by the U. S. Department of Energy, Office of Science, Basic Energy Sciences, Materials Sciences and Engineering Division. This research used resources of the Compute and Data Environment for Science (CADES) at the Oak Ridge National Laboratory, which is supported by the Office of Science of the US Department of Energy. First-principles DFT calculations were performed at the HPC of Shanghai Jiao Tong University.

## AUTHOR CONTRIBUTIONS

P.Y. and S.S. conceived the projected and designed the experiments. S.S conducted the transport measurements and analyzed all data. Z.L. and S.C. grew the samples and performed XRD measurements. Z.T., W.L. and S.O. carried out DFT+DMFT calculations. S.S. and P.Y. wrote the manuscript, and all authors commented on the paper.

# Figures and captions

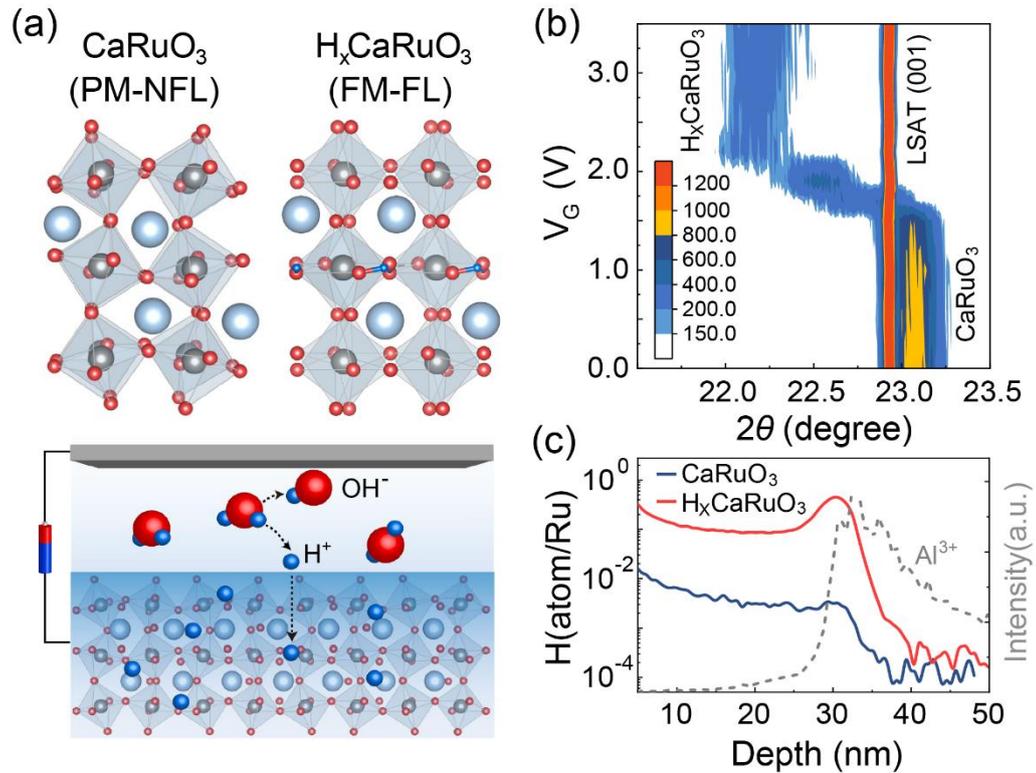

**FIG. 1. Structural transformation of CaRuO₃ through ionic liquid gating induced protonation.** (a) Schematic illustration of proton intercalated induced phase transformation in CaRuO$_3$ through ionic liquid gating (ILG). The upper panel shows a paramagnetic non-Fermi liquid (PM-NFL) metallic state in CaRuO$_3$ to ferromagnetic Fermi-liquid behavior (FM-FL) in H$_x$CaRuO$_3$ (x = 0.5). The optimized crystalline structures were obtained from DFT calculations. The lower panel shows schematic for the protonation process during the ILG. (b) *In-situ* XRD $\theta$-$2\theta$ scans around (001) peak of CaRuO$_3$ as a function of gating voltage ($V_G$). The peak position shifts gradually toward lower angle (from 23.07° to 22.17°) as the $V_G$ increases, corresponding to a distinct lattice expansion along out-of-plane for up to 4.1%. (c) Measured residual hydrogen concentration through *ex-situ* SIMS on both pristine CaRuO$_3$ and proton intercalated H$_x$CaRuO$_3$ (gated at 3.5 V) samples. A referenced Al$^{3+}$ signature was also shown to highlight the interface position between film and substrate.

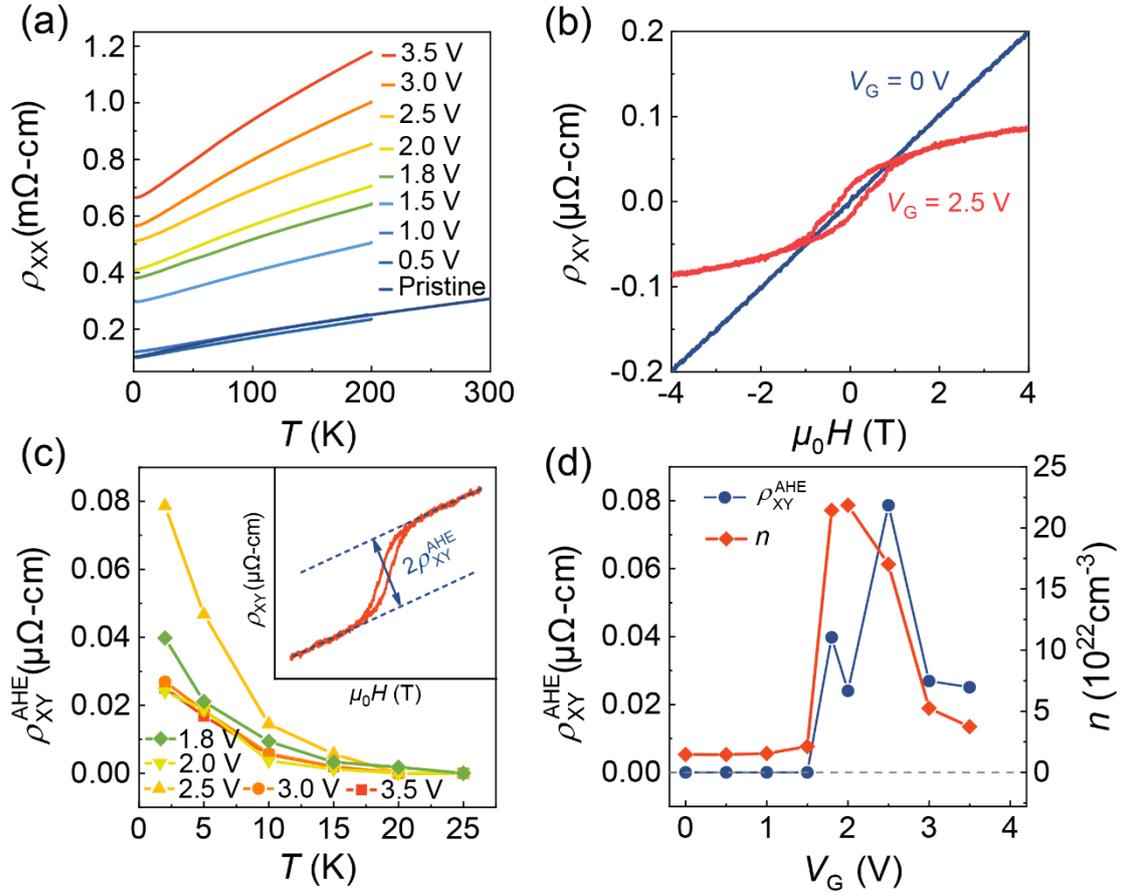

**FIG. 2. Emergent ferromagnetism in proton intercalated CaRuO3 thin films.** (a) *In-situ* temperature dependence of resistivity at different $V_G$. (b) Magnetic field dependent Hall resistivity at 2 K for pristine and gated ($V_G$ = 2.5 V) states, respectively. (c) Temperature dependent anomalous Hall resistivity $\rho_{XY}^{AHE}$ at different $V_G$. The inset illustrates the method to obtain the $\rho_{XY}^{AHE}$ from a representative Hall loop ($V_G$ = 2.0 V and $T$ = 2 K). (d) Summary of $\rho_{XY}^{AHE}$ and carrier density ($n$, extracted from linear part of Hall signal at high-magnetic-field region) as a function of $V_G$ at 2 K.

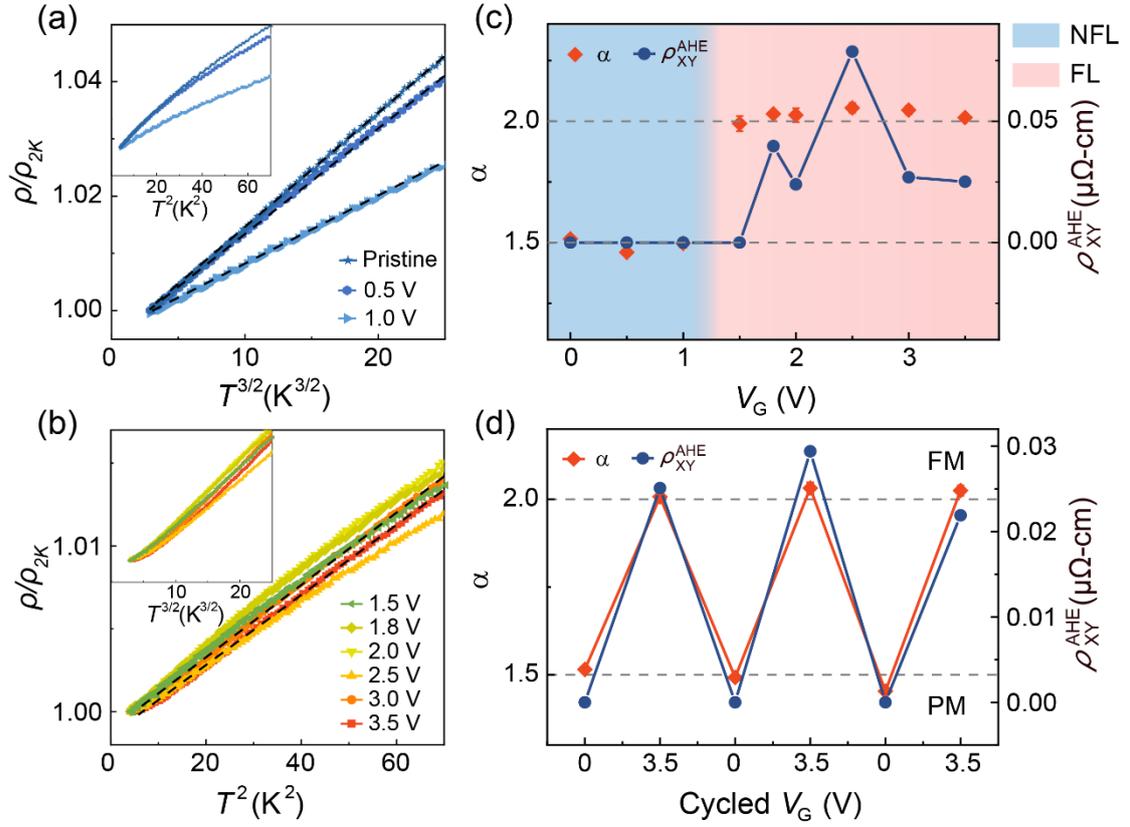

**FIG. 3. Evolution from non-Fermi liquid to Fermi liquid behavior accompanied by magnetic transition.** Normalized longitude resistivity $\rho_{xx}/\rho_{2K}$ as a function of (a) $T^{3/2}$ and (b) $T^2$ at low temperature region with different $V_G$. Insets show the deviation from $T^2$ ($T^{3/2}$) for pristine (gated at $V_G > 1.0$ V) states for comparison. The dash lines are guides to eyes. (c) Obtained fitting exponent $\alpha$ as a function of $V_G$ for $\rho(T) \sim T^\alpha$, forming two regions labelled by NFL and FL. The $V_G$ dependence of $\rho_{XY}^{AHE}$ is also shown. (d) Reversible evolution of fitting exponent $\alpha$ and $\rho_{XY}^{AHE}$, as $V_G$ cycled between 0 V and 3.5 V.

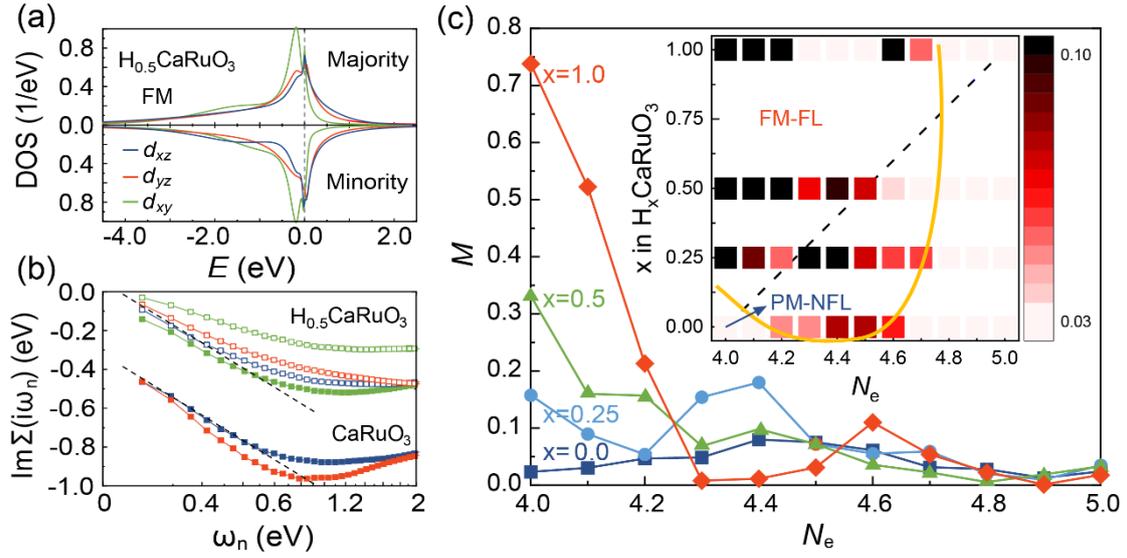

**FIG. 4. Theoretically calculated phase diagram of proton intercalated $H_xCaRuO_3$.** (a) Spin polarized density of states (DOS) of $H_{0.5}CaRuO_3$ obtained through DFT+DMFT calculations. The vertical dashed line marks the Fermi level which is taken to be the origin of the horizontal axis. (b) Imaginary part of the self-energy as a function of Matsubara frequency $\omega_n$ for $CaRuO_3$ (filled symbol) and $H_{0.5}CaRuO_3$ (open symbol). A paramagnetic state is considered for both cases. The color codes used denote different orbitals in the same way as (**a**), and the dashed lines indicate $(\omega_n)^{1/2}$ behavior. (c) Calculated Ru magnetic moment $M$ as a function of Ru $d$ electron number ($N_e$) for pristine $CaRuO_3$ (blue square), $H_{0.25}CaRuO_3$ (light-blue cycle), $H_{0.5}CaRuO_3$ (green triangle) and $HCaRuO_3$ (red diamond), respectively. The inset shows a summarized intensity map of magnetization ($M$) as a function of both $N_e$ and $x$, in which the color bar denotes $M$. The yellow line shows guideline for the boundary between ferromagnetism (FM) and paramagnetism (PM), while the dashed black line represents the actual protonation effect with simultaneous structural expansion and electron doping (as $N_e = 4 + x$).

# Supplementary information for

# Emergent ferromagnetism with Fermi-liquid behavior in proton intercalated $CaRuO_3$

Shengchun Shen[1], Zhuolu Li[1], Zijun Tian[2], Weidong Luo[2,3*], Satoshi Okamoto[4*], and Pu Yu[1,5,6*]

[1]*State Key Laboratory of Low Dimensional Quantum Physics and Department of Physics, Tsinghua University, Beijing 100084, China*
[2]*Key Laboratory of Artificial Structures and Quantum Control, School of Physics and Astronomy, Shanghai Jiao Tong University, Shanghai 200240, China*
[3]*Institute of Natural Sciences, Shanghai Jiao Tong University, Shanghai 200240, China*
[4]*Materials Science and Technology Division, Oak Ridge National Laboratory, Oak Ridge, Tennessee 37831, USA*
[5]*Frontier Science Center for Quantum Information, Beijing 100084, China*
[6]*RIKEN Center for Emergent Matter Science (CEMS), Wako 351-198, Japan*

Email: wdluo@sjtu.edu.cn; okapon@ornl.gov and yupu@mail.tsinghua.edu.cn

**Methods**

**Growth of CaRuO₃ thin films.** Thin films were deposited on (001) $(LaAlO_3)_{0.3}(SrAl_{0.5}Ta_{0.5}O_3)_{0.7}$ (LSAT) substrates through a pulsed laser deposition system, with a laser fluence of 2 J/cm² (KrF, λ = 248 nm) at the growth temperature of 700 °C and oxygen pressure of 4 Pa. To minimize the oxygen vacancy concentration, samples were annealed at growth temperature for 15 minutes after the growth and then cooled down to room temperature at a cooling rate of 10 °C per minute in an atmosphere of oxygen pressure.

***In-situ* XRD.** The crystalline structures were characterized by a high-resolution diffractometer (Smartlab, Rigaku) using monochromatic Cu Kα1 radiation (λ = 1.5406 Å) at room temperature. For *in-situ* measurements, a slice of Pt was selected as the gated electrode and Au were sputtered on the edge of films as bottom electrode. The thin films and gate electrode were then placed into a quartz bowl sitting on the sample-holder of diffractometer. Before measurements, the samples were firstly aligned with the substrate diffraction peaks, and then a small amount of ionic liquid was carefully added into the bowl to cover completely the samples, while keeping relatively large x-ray diffraction intensity for measurements. Afterward, the *θ-2θ* scans were carried out as a function of $V_G$ with the same time interval, which can then trace nicely the structural evolution through gating.

**Electrical transport measurements.** The transport measurements were performed in a PPMS setup (Quantum Design DynaCool system, 9 T) equipped with lock-in amplifiers (Model LI 5640, NF Corporation). To carry out the *in-situ* transport studies, a 60 μm × 220 μm Hall bar was fabricated through standard lithography from thin films and the electrodes were capped with sputtered Ti/Au. Subsequently, the sample was placed in a quartz bowl cover entirely with ionic liquid and a slice of Pt was used as the gate electrode. For each state, $V_G$ was changed and then held up to 20 minutes at 290 K before cooling down toward lower temperature (<240 K) to freeze the ionic liquid.

**First-principles calculations:** In density function theory (DFT) calculations, we first considered CaRuO$_3$ with its pristine orthorhombic structure (*Pnma*), which contains 8 Ru sites per structural unit cell. Then we fixed the in-plane lattice constants with those of LSAT substrate, while optimizing the out-of-plane lattice constant and atomic coordinates. We used a supercell consisting of eight chemical formula unites of CaRuO$_3$, and varied the number of hydrogen atom per cell as 0 (distorted CaRuO$_3$, *x*=0), 2 (protonated H$_{0.25}$CaRuO$_3$), 4 (protonated H$_{0.5}$CaRuO$_3$), and 8 (protonated HCaRuO$_3$). It is important to note that in proton intercalated samples, all Ru sites were distorted from their original orthorhombic coordinates, and therefore when calculating the partial density of states, we averaged over "symmetry-related" *d* orbitals in the basis which diagonalizes the local crystal field [1]. These local orbitals can then be labeled using the original notation, i.e, $d_{xz}$, $d_{yz}$ or $d_{xy}$. We further examined the ferromagnetic instability of the proton intercalated samples through the Dynamical Mean-Field Theory (DMFT) method based on the optimized crystalline structure from our DFT calculations. To reduce the computational cost, we considered one impurity model per DMFT simulation that represents the symmetrically averaged Ru sites. This DFT+DMFT then allows one to examine a wider range of electron concentration than the DFT approach alone, where the Ru *d* electron number $N_e$ and the proton concentration *x* are related as $N_e = 4 + x$.

DMFT calculations were performed using an effective model expressed as $H = H_{band} + \sum_{i \in \text{Ru } d} H_{int,i}$. The band part $H_{band}$ is expressed in the Wannier basis, and the interaction part $H_{int,i}$ is given by

$$H_{int,i} = U \sum_\alpha d^\dagger_{i\alpha\uparrow} d_{i\alpha\uparrow} d^\dagger_{i\alpha\downarrow} d_{i\alpha\downarrow} + U' \sum_{\alpha \neq \beta} d^\dagger_{i\alpha\uparrow} d_{i\alpha\uparrow} d^\dagger_{i\beta\downarrow} d_{i\beta\downarrow}$$

$$+(U' - J_H) \sum_{\alpha>\beta,\sigma} d^\dagger_{i\alpha\sigma} d_{i\alpha\sigma} d^\dagger_{i\beta\sigma} d_{i\beta\sigma} + J_H \sum_{\alpha \neq \beta} (d^\dagger_{i\alpha\uparrow} d_{i\beta\uparrow} d^\dagger_{i\beta\downarrow} d_{i\alpha\downarrow} +$$

$$d^\dagger_{i\alpha\uparrow} d_{i\beta\uparrow} d^\dagger_{i\alpha\downarrow} d_{i\beta\downarrow}).$$

Here, $d^{(\dagger)}_{i\alpha\sigma}$ is the annihilation (creation) operator for a Ru *d* (t$_{2g}$) electron at site *i*, orbital $\alpha$ (=$d_{xz}$, $d_{yz}$ or $d_{xy}$), with spin $\sigma$ (↑ or ↓). $U$ is the intra-orbital Coulomb interaction, $U'$ is the inter-orbital Coulomb interaction, and $J_H$ is the inter-orbital

Hund coupling. We took $U = 2.3$, $J_H = 0.4$ (eV) with $U' = U - 2J_H$ from Ref. 1.

The DMFT procedure can then map this interacting lattice model into interacting impurity models consisting of interacting Ru $d$ states and the effective hybridization that will be solved self-consistently[2]. In principle, 8 Ru sites are inequivalent in protonated CaRuO$_3$. However, these Ru sites are approximately related by the orthorhombic symmetry. Thus, we diagonalized the local part of $H_{band}$, and averaged their effective hybridization functions over "symmetry-related" orbitals. This operation reduces the number of impurity models from 8 to 1 (Ref: 1). Further, we kept only diagonal components of the hybridization functions, by which the negative sign problem could be eliminated. To solve the impurity model, we used the hybridization-expansion version of the continuous-time quantum Monte-Carlo method[3] at $T = 0.01$ eV = 116 K. The quantum Monte-Carlo procedure necessarily worked on the imaginary-time or the imaginary-frequency axis, thus we analytically continued electron Green's functions from the imaginary frequency axis to the real frequency axis using the maximum entropy method[4].

**Supplemental Figures**

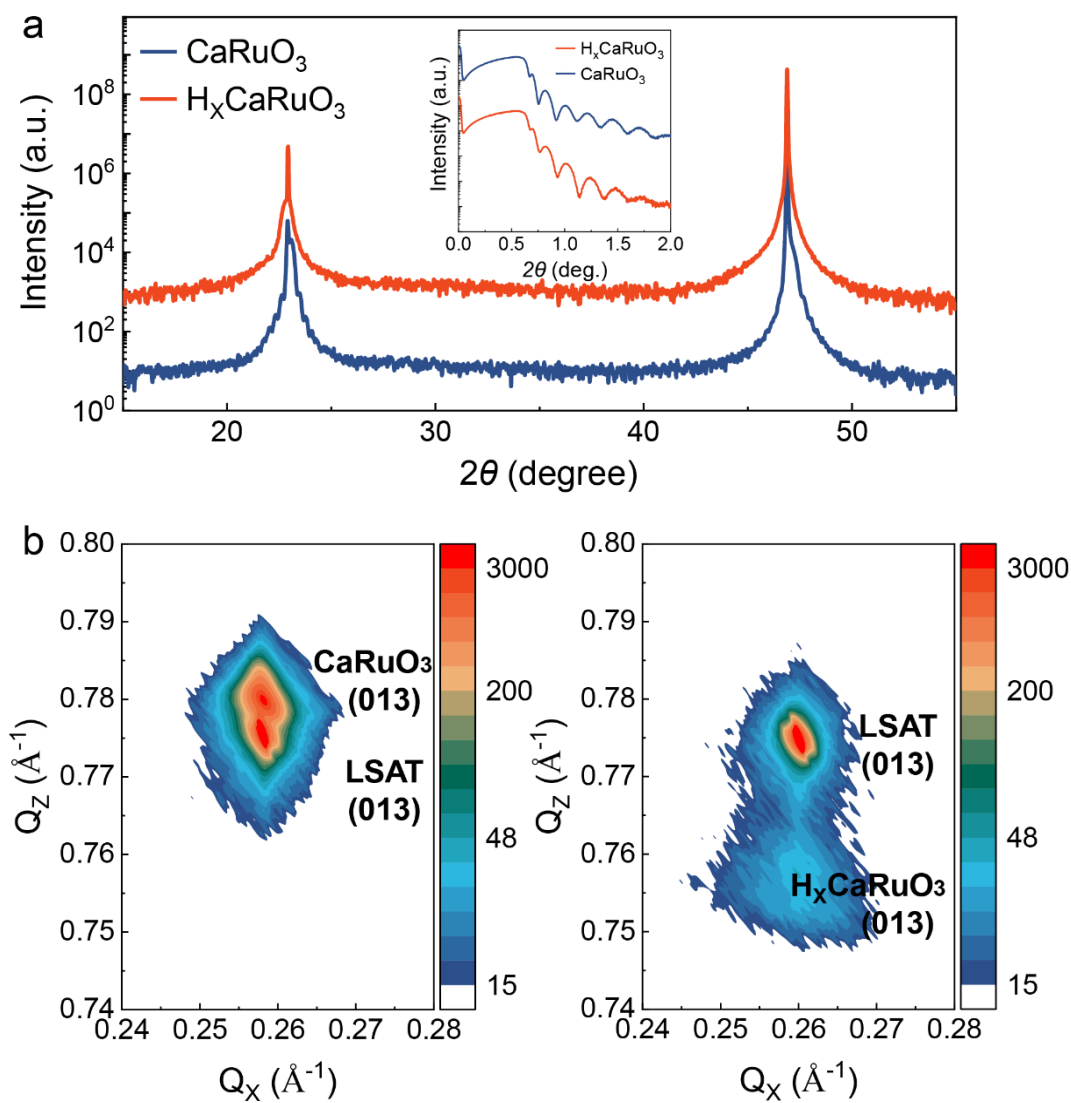

**Supplemental Figure 1.** Structural characterizations of pristine $CaRuO_3$ and proton intercalated $H_XCaRuO_3$ (gated at $V_G$ = 3.5 V) films. (a) Extended-range XRD $\theta$-$2\theta$ scans of $CaRuO_3$ thin film (31 nm) at pristine and gated (at $V_G$ = 3.5 V, *ex-situ*) states. The inset shows the reflectometry data for both pristine and protonated states, in which the sample roughness remains almost unchanged as evidenced by the clear thickness fringes. (b) *In-situ* reciprocal space mappings (RSM) around (013) peak of LSAT substrate for both pristine and proton intercalated states. A thicker film with thickness of 90 nm was used during the RSM measurements to obtain large intensity, which however remains epitaxially coherent with the substrate due to the associated small lattice mismatch.

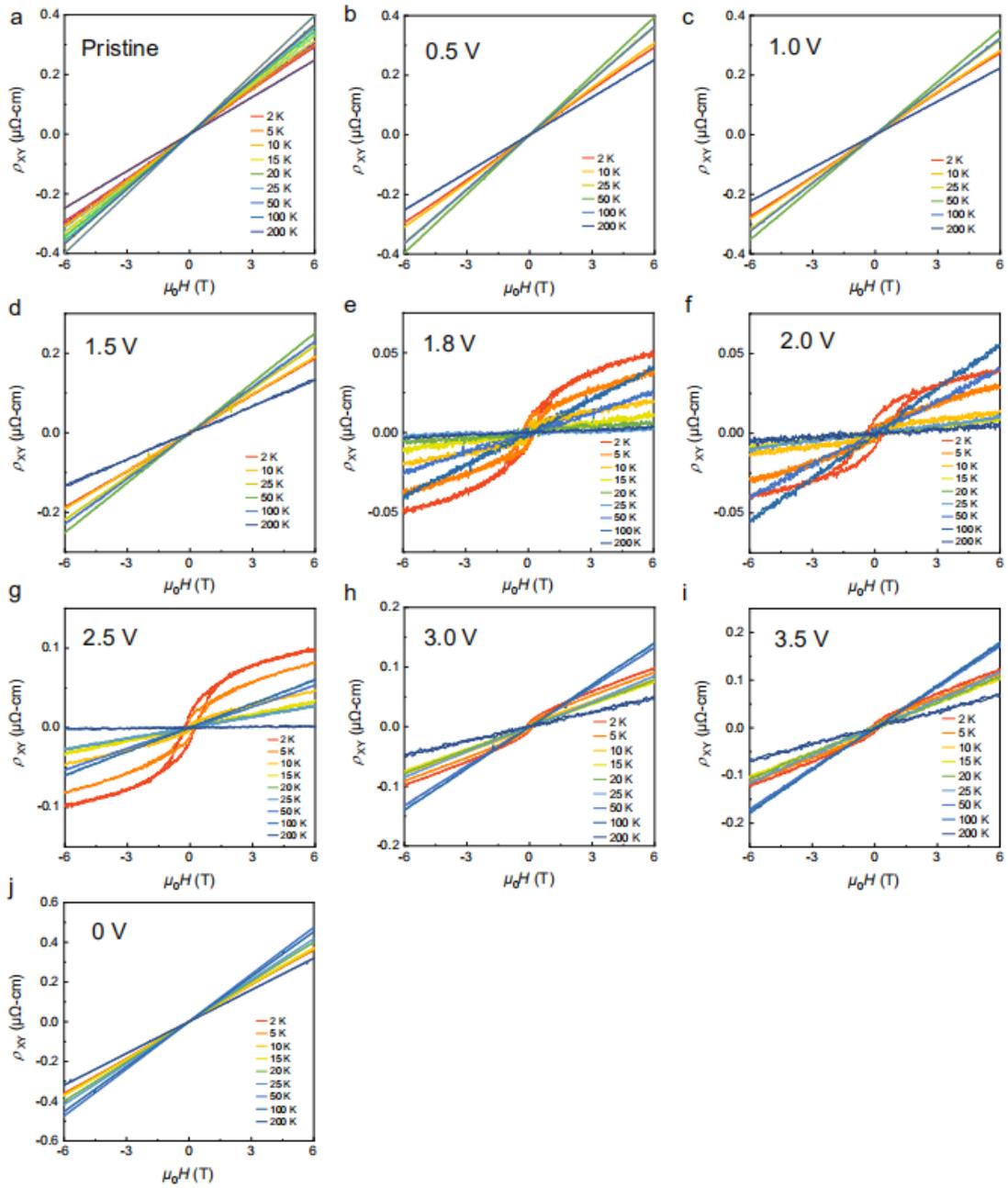

**Supplemental Figure 2.** Extended magnetic-field dependent Hall resistivity in CaRuO$_3$ during ionic liquid gating. The voltage was gradually increased from 0 V (a), to 0.5 V (b), to 1.0 V (c), to 1.5 V (d), to 1.8 V (e), to 2.0 V (f), to 2.5 V (g), to 3.0 V (h), to 3.5 V (i), and then returned back to 0 V (j).

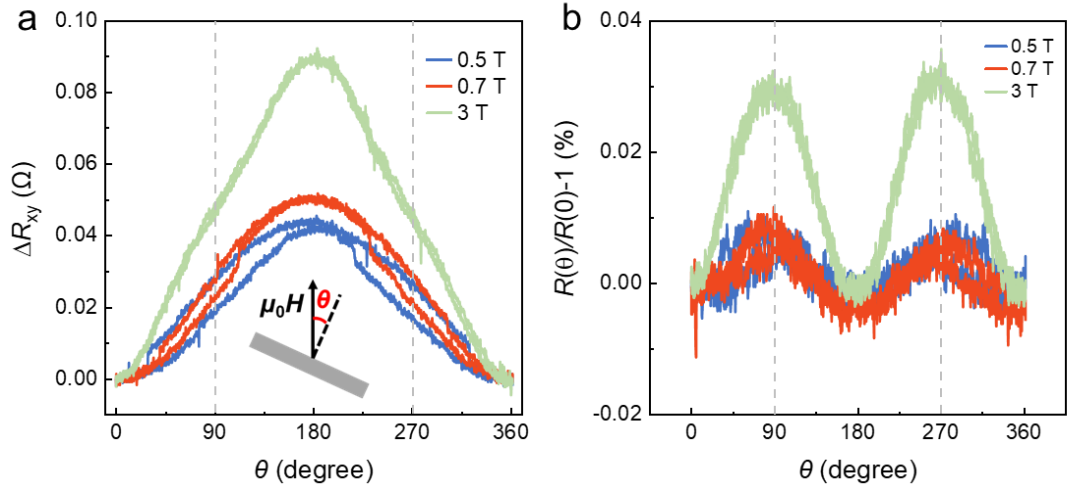

**Supplemental Figure 3.** Magnetic field angle dependence of (b) Hall resistivity ($R_{xy}(\theta) - R_{xy}(0)$) and (c) longitudinal resistivity ($R(\theta)/R(0)-1$) for ferromagnetic HxCaRuO3 measured at 2 K with different magnetic field. $\theta$ denotes the angle between magnetic field and the normal of film. The reversal of the magnetic moment leads to jumps in Hall resistance and peaks in longitudinal resistance around 90° and 270° at 0.5 T or 0.7 T. And under a larger applied magnetic field (3 T), the magnetic moment follows the changes in the direction of field, the jumps and peaks would be suppressed. Above observations suggest the magnetic easy axis in ferromagnetic $H_xCaRuO_3$ is along out-of-plane direction.

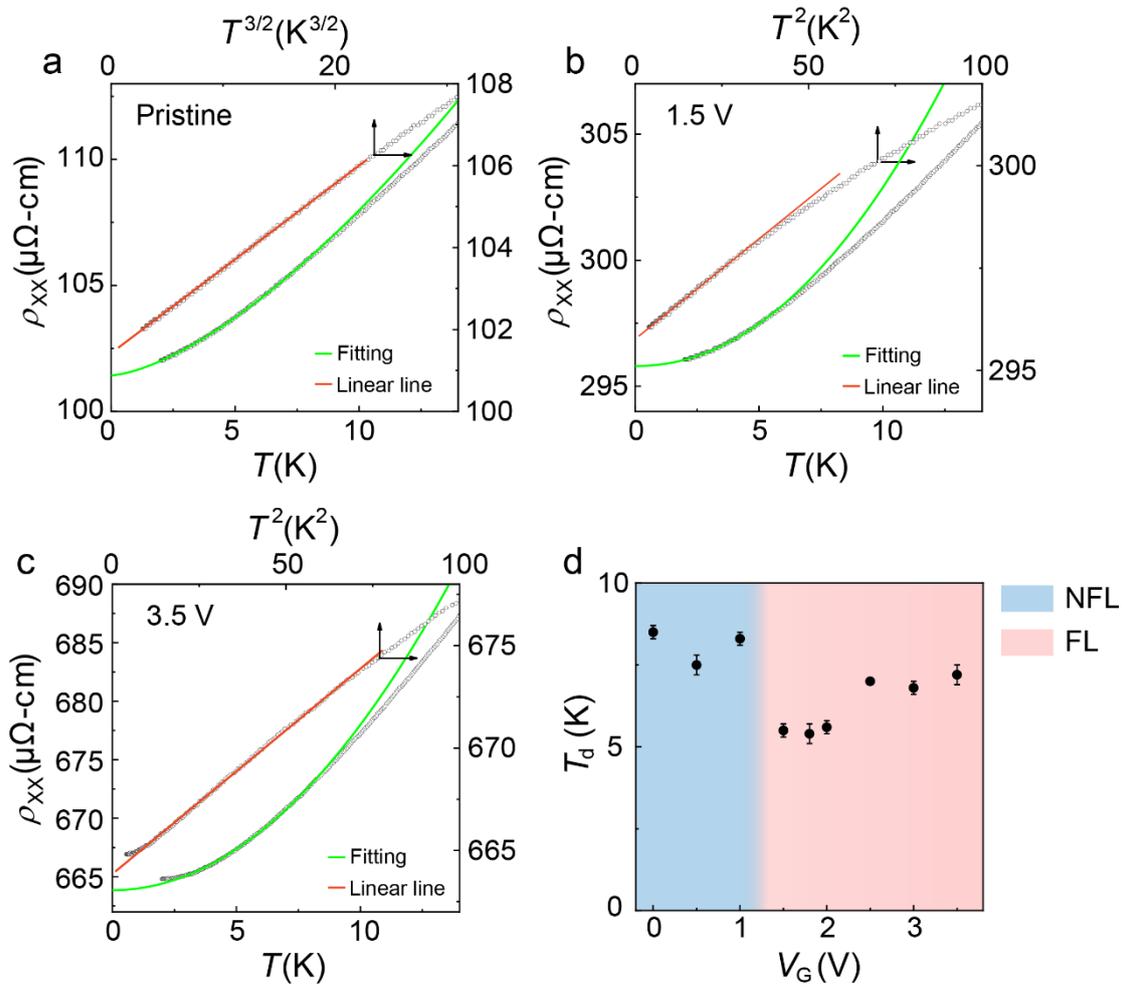

**Supplemental Figure 4.** Direct comparison of $\rho_{xx}(T)$ curves for (a) pristine state, and gated at (b) 1.5 V and (c) 3.5 V. The red lines show the linear fitting with $\rho_{xx}(T) \sim T^{3/2}$ (or $T^2$) relationship at low temperature region. The green lines show corresponding empirical fittings with the formula of $\rho = \rho_0 + AT^\alpha$ at low temperature region. The fitting parameters were extracted with the linear correlation coefficient R-square > 0.998. (d) The $V_G$ dependence of characterized temperature ($T_d$) at which the $\rho_{xx}(T)$ curve deviates from the $\rho_{xx}(T) \sim T^{3/2}$ (or $T^2$) relationship.

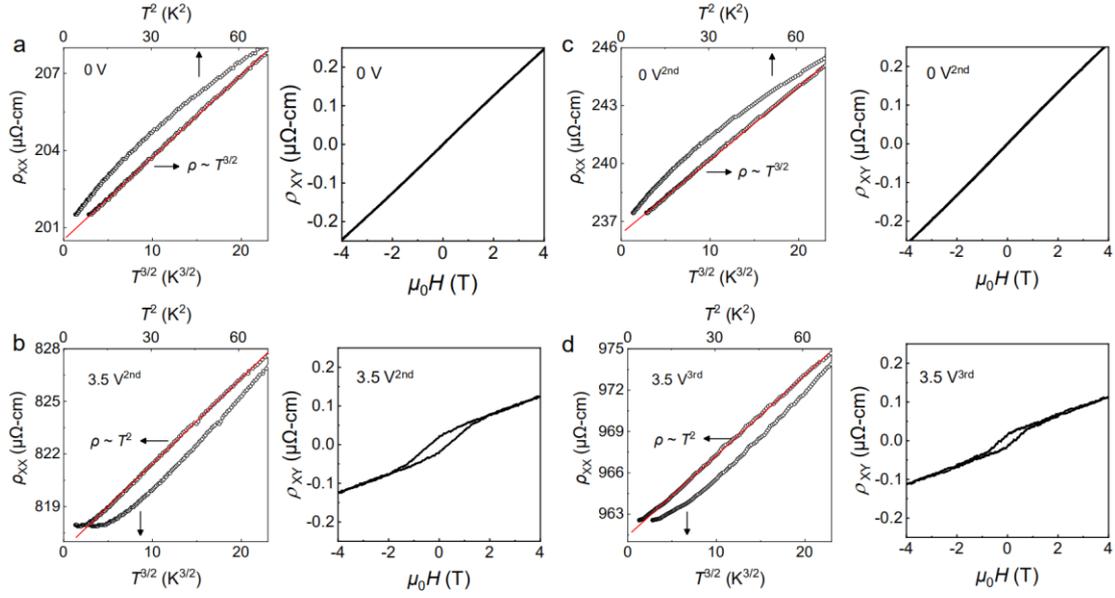

**Supplemental Figure 5.** Reversibility test of the paramagnetic to ferromagnetic transition. Temperature dependent resistivity and magnetic field dependent Hall resistivity (at 2 K) as $V_G$ cycled between (a, c) 0 V and (b, d) 3.5 V along the sequence of a-b-c-d.

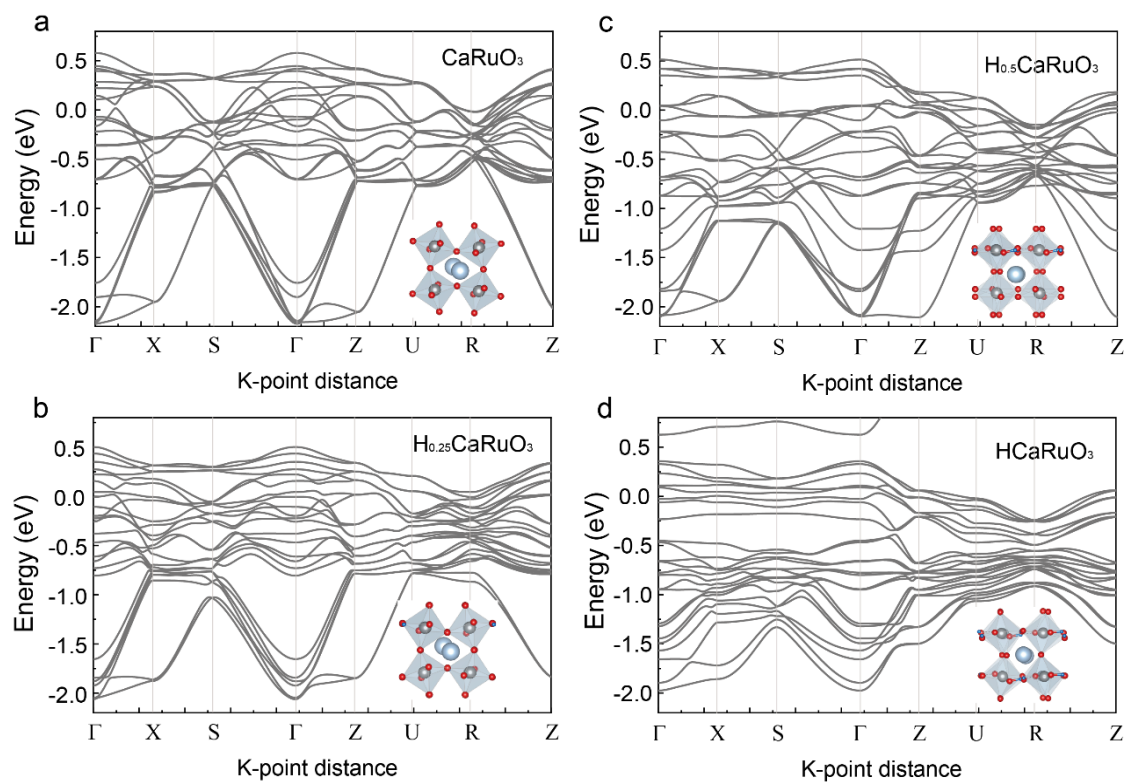

**Supplemental Figure 6.** Calculated non-spin-polarized Ru $t_{2g}$ band structures for (a) pristine CaRuO$_3$, (b) H$_{0.25}$CaRuO$_3$, (c) H$_{0.5}$CaRuO$_3$, and (d) HCaRuO$_3$ obtained from DFT calculations with nonmagnetic General Gradient Approximation (GGA). The inset figures show the corresponding calculated lattice structures. The calculated Ru-O-Ru bonding angles are 148.1 deg. for CaRuO$_3$, 149.5 deg. for H$_{0.25}$CaRuO$_3$, 156.1 deg. for H$_{0.5}$CaRuO$_3$ and 170.8 for HCaRuO$_3$, respectively.

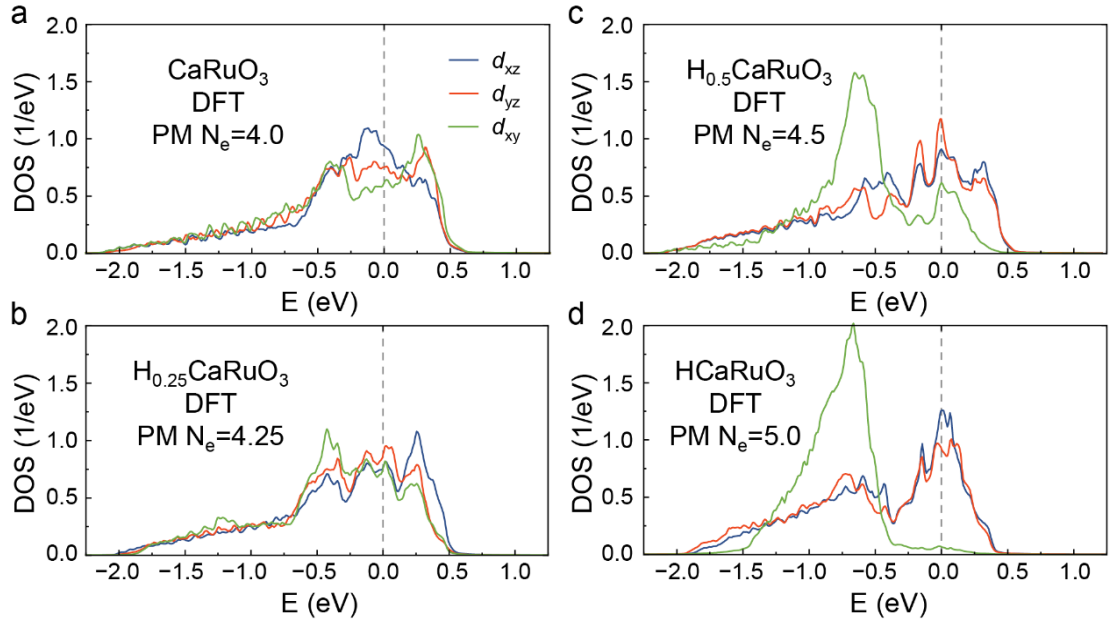

**Supplemental Figure 7.** Orbital-resolved density of states (DOS) for (a) $CaRuO_3$, (b) $H_{0.25}CaRuO_3$, (c) $H_{0.5}CaRuO_3$, and (d) $HCaRuO_3$ obtained from DFT calculations with paramagnetic ground state. For pristine and $H_{0.25}CaRuO_3$ structures (a, b), all three orbitals are almost degenerate and the peak positions occur around the Fermi level, which is taken to be the origin of the horizontal axis, at relatively low density. While for larger hydrogen concentrations (c, d), the DOS shows a broad peak for the $d_{xy}$ band located far below the Fermi level, while that for the $d_{xz}/d_{yz}$ orbitals develop a sharp peak at $E=0$. Such an orbital reconstruction is indeed consistent with the scenario of chemical intercalation induced lattice expansion along $c$ axis.

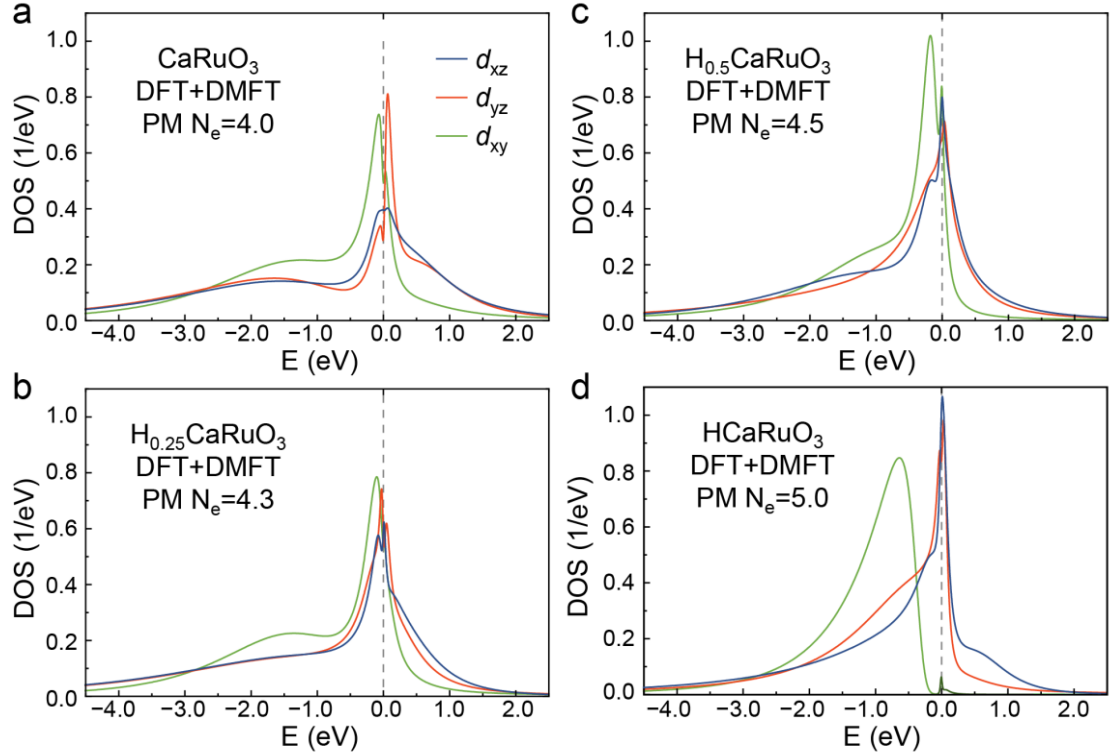

**Supplemental Figure 8.** Orbital-resolved density of states (DOS) for (a) $CaRuO_3$, (b) $H_{0.25}CaRuO_3$ ($N_e$ = 4.3), (c) $H_{0.5}CaRuO_3$, and (d) $HCaRuO_3$ obtained from DFT + DMFT calculations with paramagnetic state. The proton intercalation process leads to notable orbital reconstructions with enhanced DOS for all three orbitals and a shift of $d_{xy}$ level toward lower energy from the Fermi level, which is taken to be the origin of the horizontal axis. The DFT+DMFT calculations for pristine $CaRuO_3$ are consistent with the results in literature[1].

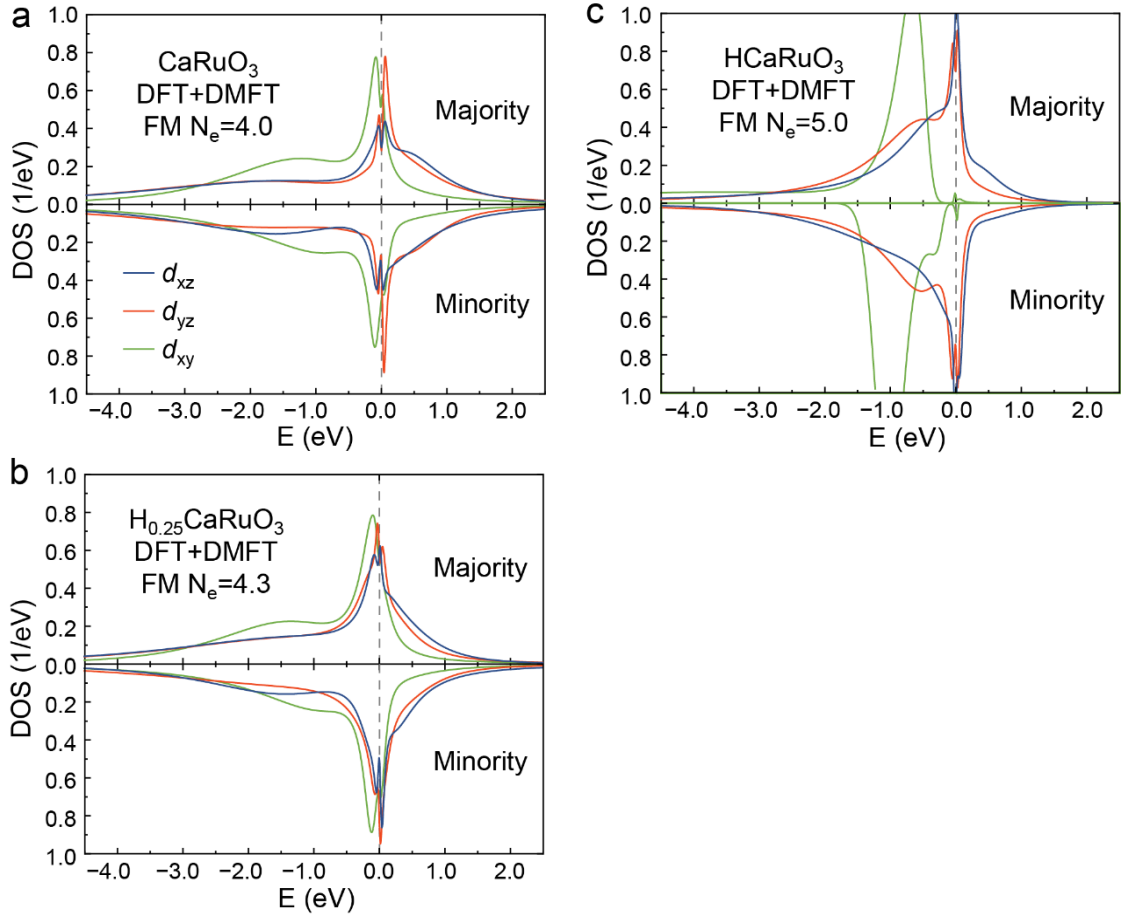

**Supplemental Figure 9.** Evolution of magnetic states through proton intercalation. Spin polarized DOS for (a) pristine CaRuO$_3$ with $N_e$ = 4.0 (b) H$_{0.25}$CaRuO$_3$ with $N_e$ = 4.3 and (d) HCaRuO$_3$ with $N_e$ = 5.0, respectively. The existence of difference between majority and minority DOS at Fermi level clearly suggests the emergent ferromagnetism at H$_{0.25}$CaRuO$_3$ with $N_e$ = 4.3, while nearly the same majority and minority DOS at Fermi level denotes the ferromagnetism is absent in CaRuO$_3$ ($N_e$ = 4.0) and HCaRuO$_3$ ($N_e$ = 5.0).

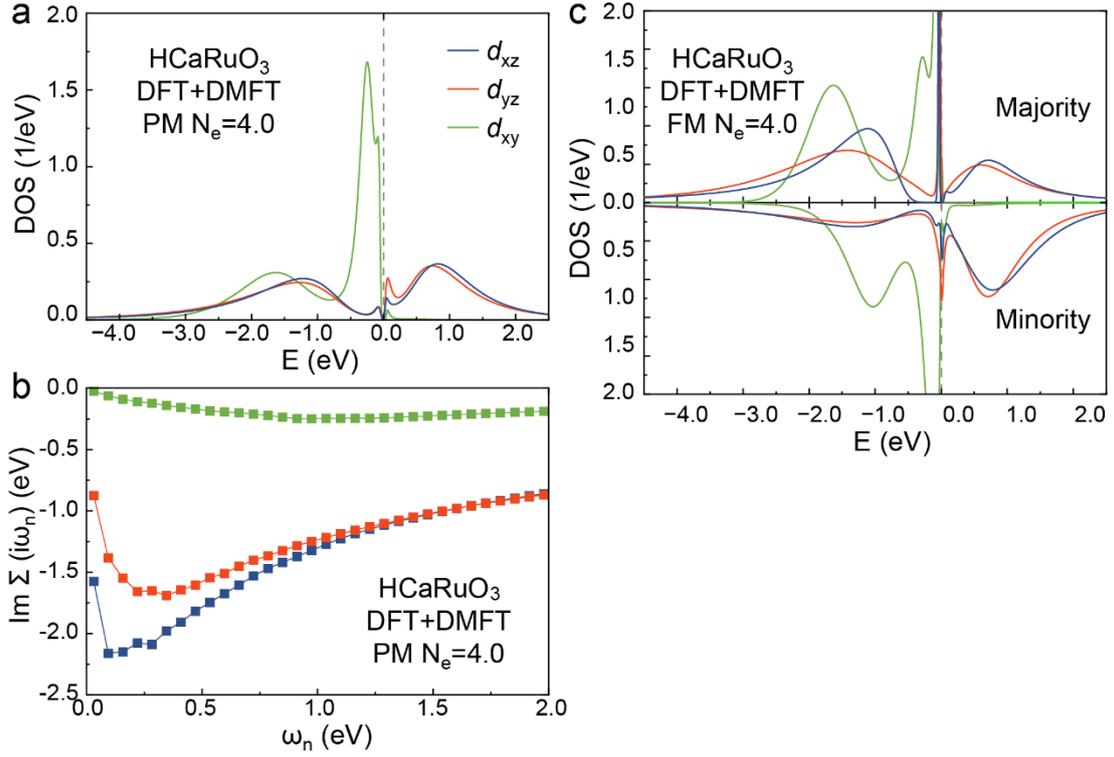

**Supplemental Figure 10.** Correlation effects in HCaRuO$_3$ with $N_e = 4.0$ obtained from DFT+DMFT calculations. (a) Orbitally resolved DOS and (b) Imaginary part of self-energy as a function of Matsubara frequency $\omega_n$ for a paramagnetic state. As indicated by a small but finite gap in DOS for $d_{xz}$ and $d_{yz}$ bands and the strongly enhanced self-energy, this result shows an orbital-selective Mott insulating state. (c) Spin polarized DOS for a ferromagnetic state. Large difference between majority and minority DOS at Fermi level denotes a strong ferromagnetism.